 \documentclass{ecai} 

\usepackage{latexsym}
\usepackage{amssymb}
\usepackage{amsmath}
\usepackage{amsthm}
\usepackage{booktabs}
\usepackage{enumitem}
\usepackage{graphicx}
\usepackage{color}
\usepackage{algorithm}
\usepackage{algorithmicx}
\usepackage{algpseudocode}
\usepackage[switch]{lineno}
\usepackage{tikz}
\usepackage{booktabs}
\usepackage{mathtools}
\usepackage{amssymb}
\usepackage{multirow}
\usepackage{multicol}
\usepackage{subcaption}
\usepackage{booktabs}
\usepackage{stfloats}

\DeclareMathOperator{\Universe}{\mathcal{U}}

\newtheorem{theorem}{Theorem}
\newtheorem{lemma}[theorem]{Lemma}

\newtheorem*{definition*}{Definition}

\newcommand{\BibTeX}{B\kern-.05em{\sc i\kern-.025em b}\kern-.08em\TeX}

\begin{document}


\begin{frontmatter}


\paperid{7351} 


\title{Efficient Branch-and-Bound for Submodular Function Maximization under Knapsack Constraint}




\author[A]
{\fnms{Yimin}~\snm{Hao}}

\author[A]{\fnms{Yi}~\snm{Zhou}\orcid{0000-0002-9023-4374}\thanks{Corresponding Author. Email: zhou.yi@uestc.edu.cn}}

\author[A]{\fnms{Chao}~\snm{Xu}\orcid{0000-0003-4417-3299}} 

\author[B]{\fnms{Zhang-Hua}~\snm{Fu}\orcid{0000-0002-3740-7408}}

\address[A]{University of Electronic Science and Technology of China, China}
\address[B]{Shenzhen Institute of Artificial Intelligence and Robotics for Society, The Chinese University of Hong Kong, Shenzhen, China}


\begin{abstract}
The submodular knapsack problem (SKP), which seeks to maximize a submodular set function by selecting a subset of elements within a given budget, is an important discrete optimization problem. 
The majority of existing approaches to solving the SKP are approximation algorithms.
However, in domains such as health-care facility location and risk management, the need for optimal solutions is still critical, necessitating the use of exact algorithms over approximation methods. 
In this paper, we present an optimal branch-and-bound approach, featuring a novel upper bound with a worst-case tightness guarantee and an efficient dual branching method to minimize repeat computations.
Experiments in applications such as facility location, weighted coverage, influence maximization, and so on show that the algorithms that implement the new ideas are far more efficient than conventional methods. 
\end{abstract}

\end{frontmatter}


\section{Introduction}

A function $f:2^\mathcal{U}\rightarrow \mathbb{R}$ for a universe $\Universe$ is called a \textit{set function}.
Given a set function $f$, a subset $S \subseteq \mathcal{U}$ and an element $e \in \mathcal{U}$, define $f(e|S)$ as the \textit{marginal gain} of $f$ with respect to $e$, i.e. $f(e|S) = f(S\cup \{e\})-f(S)$.
Then, a set function $f$  is \textit{submodular} if for any $S\subseteq T \subseteq \Universe$ and $e\in \Universe \setminus T$, $f(e|S)\ge f(e|T)$ and \textit{monotone} if for any $S \subseteq \Universe$ and $e \in \Universe$, $f(e|S) \geq 0$.
In other words, a submodular function is a set function with diminishing marginal gains.
In the paper, we consider the well-known monotone submodular maximization problem under a \textit{knapsack constraint}, which we refer to as the \textit{submodular knapsack problem} (SKP). Given a monotone submodular function $f$ and a budget $W\in \mathbb{R}_+$, supposing each element $e \in \mathcal{U}$ is assigned a cost value $w_e$, the SKP seeks a subset $S\subseteq \mathcal{U}$ such that the total cost of $\sum_{e\in S}w_e\le W$ is at most $W$ but $f(S)$ is maximized.

The SKP has been widely used to model the decision problems in the applications where the law of diminishing returns property holds.
One such example is categorical feature compression,  which can be modeled as the problem of maximizing mutual information between the compressed version of the categorical features and the target label \cite{bateni2019categorical}. 
Furthermore, problems such as the maximum weighted coverage problem \cite{khuller1999budgeted} and the influence maximization problem \cite{kempe2003maximizing,alon2012optimizing} can also be expressed as the SKP problems.
Recent applications concerning the SKP also emerged in machine learning \cite{bach2013learning}, document summarization \cite{lin2010multi,lin2011class}, and sensor placement \cite{krause2008near}.

In terms of the hardness of the SKP, since the Hitting Set problem can be reduced to the SKP, there exists no algorithm with a time complexity of $O((2-\epsilon)^{|\Universe|})$ unless Strong Exponential Time Hypothesis fails \cite[Thm. 14.35.]{cygan2015parameterized}.
Several greedy algorithms are often used to find solutions with approximation guarantees to the optimal.
Specifically, the greedy algorithm \cite{lin2010multi} can obtain an $(1-1/\sqrt{e})$-approximation solution in $O(|\Universe|^2T(f))$ time, where $T(f)$ denotes the running time of the oracle $f(S)$. A $(1 - 1/e)$-approximation solution can be obtained by the greedy algorithm of running time $O(|\Universe|^5T(f))$ \cite{sviridenko2004note}. 
However, the SKP cannot be approximated within $(1 - 1/e + \epsilon)$ ($\epsilon>0$) in polynomial time unless $P=NP$ \cite{feige1998threshold}.
From the application perspective, in order to handle large datasets or to accommodate new machine architectures, approximate parallel algorithms \cite{balkanski2019exponential} and streaming algorithms \cite{huang2021improved} are also studied for the past several years.

Nevertheless, in some optimality-critical applications, an approximate solution may not be sufficient.
For example, in health-care facility location \cite{ahmadi2017survey}, facilities may be used for many years once decided, which allows one to pursue the optimal location at the initial decision phase.
In risk decision-making \cite{wilder2018risk}, where decisions are made based on the conditional value at risk (CVaR), an approximate solution may lead to substantial economic losses, which allows us to use a higher computational cost for trading the optimal solutions. 

Overall, there is still a significant demand for exact algorithms to resolve the SKP.
As far as we know, \citeauthor{nemhauser1981maximizing} proposed the first exact algorithm which is based on integer linear programming \cite{nemhauser1981maximizing}.
Later, an A$^*$ algorithm, which can produce near-optimal solutions of a given approximation ratio $\alpha$, was proposed in \cite{chen2015filtered}. 
By setting the hyperparameter $\alpha$ as 1, this $A^*$ algorithm becomes an exact algorithm with a sacrifice in running time.
Subsequently, this $A^*$ algorithm was further improved in \cite{sakaue2018accelerated} by providing a new upper bound, resulting in an empirically more efficient $A^*$ algorithm.

If the weight of each element in $\Universe$ is the same, the SKP becomes the special \textit{submodular function maximization problem under the cardinality constraint}. 
The problem has been well studied for the past decades. Particularly, in terms of the exact algorithm, the integer linear formulation of \citeauthor{nemhauser1981maximizing} was refined in \cite{kawahara2009submodularity}, \cite{uematsu2020efficient} and \cite{csokas2024constraint} for solving this special case of SKP. 
More recently, \citeauthor{woydt2024submodst} \cite{woydt2024submodst} showed the pure combinatorial branch-and-bound algorithm and achieved the state-of-the-art performance for solving this cardinality constrained problem. They used techniques like lazy updates and advanced heuristics in the branching algorithm.

In this paper, we study the branch-and-bound algorithm for exactly solving the SKP. 
We mainly investigate critical components including upper bound estimation and branching methods in a branch-and-bound framework.
Notably, we highlight the following novel elements.

\begin{enumerate}

    \item \textbf{A new refined subset upper bound estimation algorithm.}    
    The new bounding algorithm runs in time $O(|\Universe|^2(T(f) + \log |\Universe|))$. It brings small computational overhead because it can be computed simultaneously with a greedy primal heuristic during branch-and-bound.
    When the weight $w_i$ of each element is sufficiently small relative to the budget $W$, this new bound does not exceed $\frac{1}{1-e^{-1}} \approx 1.582$ times the optimal solution. 
    We also show that the bound is always at least as tight as the intuitive \textit{fractional knapsack bound}.
    And our refined subset bound is tighter than the existing upper bounds empirically.
        
    \item \textbf{A dual branching method to reduce repeat computations.}
    In a basic branch-and-bound framework, we need to run the primal heuristic GreedyAdd at every node from scratch.
    We propose a novel dual branching method that saves some repeat computation of the primal heuristic.
    We showed that the dual branching method speeds up the algorithm on average 2 times over a basic branching method in experiments.
    
    
\end{enumerate}

Other speedup techniques, such as lazy updates and several reduction methods, are also incorporated into the algorithm.
Experiments show that our algorithms that implement new ideas are significantly faster than the existing exact algorithms. 
Source codes and data are publicly available \url{https://github.com/Chhokmah0/submodKC}.
 


\section{Notations}
Assume that there is a positive weight $w_e\in \mathbb{R}_{+}$ for each element $e\in \Universe$. 
Let $w(S)$ denote $\sum_{e \in S} w_e$.
Given a budget $W \in \mathbb{R}_{+}$, we say that a set $S\subseteq \Universe$ satisfies the knapsack constraint if $w(S)\le W$.
Given a monotone submodular function $f$ and a budget $W$, we consider the \textit{submodular knapsack problem} (SKP), i.e. $\max_{S\subseteq \Universe} f(S)\ s.t.\ w(S) \le W$.
For simplicity, we also use $\text{SKP}(f(\cdot), \Universe, W)$ to denote the problem.

Similar to the definition of $f(e|S)$ at the beginning of the paper, given a set $X\subseteq \mathcal{U}$, we further denote $f(X|S)= f(X\cup S) - f(S)$. 
It is straightforward that if $f(X)$ is a monotone submodular function with respect to (w.r.t.) $X$, then given a set $S$, $f(X|S)$ is also a monotone submodular function w.r.t. $X$. 

Note that there is also another equivalent definition of the submodular set function:
a set function $f$ is a submodular set function if for any $A,B\subseteq \Universe$, $f(A)+f(B)\ge f(A\cup B) +f(A\cap B)$. 


\section{The General Scheme of Branch-and-Bound}
The basic branch-and-bound scheme is described in Alg. \ref{alg:branchbound}.

It is well-known that the branch-and-bound follows the tree search paradigm.
We denote a search node of the tree as $T=(S_T, C_T, W_T)$, where $S_T\subseteq \Universe$ represents the selected set,  $C_T \subseteq \mathcal{U}$ represents the candidate set which is disjoint from $S_T$, and $W_T$ is the remaining budget at the current node $T$, i.e. $W_T= W-w(S_T)$. 
At node $T$, our aim is to solve the \textit{restricted SKP} w.r.t. $T$,
\[
\max_{S_T\subseteq S \subseteq S_T\cup C_T} f(S),s.t.\ w(S) \leq W
\],
or equally,
\[
f(S_T)+\max_{X\subseteq C_T} f(X|S_T),s.t.\ w(X) \leq W_T
\]
which is denoted as $\text{SKP}(f(\cdot|S_T)+f(S_T), C_T, W_T)$.
In Alg. \ref{alg:branchbound}, the \textit{BranchBound} procedure solves this restricted optimization problem in a depth-first manner.

Clearly, $T=(\emptyset, \mathcal{U}, W)$ is the root node of the whole branch-and-bound tree. Solving the restricted SKP at this root node $T$ is equal to solving the original problem. 
Therefore, in Alg. \ref{alg:branchbound}, the whole algorithm is started by calling BranchBound with $T=(\emptyset, \mathcal{U}, W)$. We maintain $lb$ and $S^*$ as the best-known lower bound and the solution set in the tree search.

\begin{algorithm}[!t]
    \caption{The branch-and-bound for the SKP} \label{alg:branchbound}
    \textbf{Input}: The universe $\Universe$, weight vector $w_e$ for any $e\in U$, oracle to $f(\cdot)$, budge $W$.\\
    \textbf{Output}: The maximum objective $lb^*$, the best subset $S^*$ of maximum objective.
    \begin{algorithmic}[1] 
        \Function{SKP}{$\Universe, w, f(\cdot),W$}
            \State $lb^* \gets 0$ \Comment{Global Variable}
            \State $S^* \gets \emptyset$ \Comment{Global Variable}
            \State BranchBound($T=(\emptyset, \Universe, W)$)
            \State \textbf{return} $(lb^*, S^*)$
        \EndFunction 
        \Function{BranchBound}{$T=(S_T,C_T,W_T)$}
            \State $S'\gets $ a solution by primal heuristic 
            \If{$f(S') > lb^*$}
                \State $lb^* \gets f(S')$, $S^*\gets S'$
            \EndIf
            \State $ub \gets$ estimate an upper bound at node $T$
            \If{$ub \leq lb^*$}
                \State \Return
            \EndIf
            \State Branching and obtain child nodes $T_1,...,T_b$
            \For{$T \gets T_1,...,T_b$}
                \State BranchBound($T_i$)
            \EndFor
        \EndFunction
    \end{algorithmic}
\end{algorithm}

\begin{algorithm}[!t]
    \caption{The process of adding elements greedily.} \label{alg:greedy_add}
    \textbf{Input}: The search node $T = (S_T, C_T,W_T)$, oracle to $f(\cdot)$.\\
    \textbf{Output}: The greedy solution $S'$
    \begin{algorithmic}[1] 
        \Function{GreedyAdd}{$T = (S_T, C_T,W_T)$}
            \State $X \gets \emptyset$
            \State $C \gets C_T$
            \While{$C$ is not empty}
                \State $v \gets \arg\max_{v \in C}\frac{f(v|S_T\cup X)}{w_v}$
                \If{$w(X \cup \{v\}) \leq W_T$}
                    \State $X \gets X \cup \{v\}$
                \EndIf
                \State $C \gets C \setminus \{v\}$
            \EndWhile
            \State \Return $S_T \cup X$
        \EndFunction
    \end{algorithmic}
\end{algorithm}



\paragraph{The Greedy Primal Heuristic}
In the branch-and-bound, a primal heuristic algorithm quickly produces a feasible solution at each node (see line 8 in Alg. \ref{alg:branchbound}).
We introduce a simple GreedyAdd algorithm in Alg. \ref{alg:greedy_add} as a basic primal heuristic.

Given a set $S\subseteq \Universe$ and an element $e\in \Universe$, let us define the \textit{unit gain of $e$} with respect to $S$ as $f(e|S)/w_e$. 
In general, GreedyAdd iteratively expands the set $X$ by adding the element $v \in C_T$ with the highest unit gain, and ensures the addition of $v$ does not violate the knapsack constraint.
The running time of this primal heuristic is $O(|C_T|^2T(f))$.

As a matter of fact, GreedyAdd is a part of the existing $(1 - 1/\sqrt{e})$-approximation algorithm in \cite{lin2010multi}. The approximation algorithm additionally asks for a comparison of $X$ with a singleton subset of $C_T$ with maximum marginal gain. 
We select GreedyAdd as the primal heuristic for two reasons: (1) the additional step conflicts with the lazy update technique (Section 6), and (2) the empirical results show no performance improvement when incorporating this step.

\section{Upper Bound Estimation}
Given a node $T = (S_T, C_T, W_T)$, we investigate the methods to estimate an upper bound of the restricted SKP w.r.t. $T$, i.e. au upper bound to SKP$(f(\cdot|S_T)+f(S_T), C_T, W_T)$. If the upper bound is smaller than $lb$, then the search with $T$ can be pruned.


\paragraph{Domination Bound} 
A simple idea for bounding is to make use of existing approximation algorithms. 
Given a node $T = (S_T, C_T, W_T)$, supposing there is an $\alpha$-approximation algorithm that can obtain a solution $S' \subseteq C_T$, then $f(S_T)+f(S'|S_T)/\alpha$ is a valid upper bound for this node $T$. 

Indeed, this idea has been extensively exploited by \citeauthor{sakaue2018accelerated} in \cite{sakaue2018accelerated}. 
They used the approximation algorithm in \cite{lin2010multi} and derived a \textit{domination bound}.
We denote their bound as $ub_{dom}(T)$ in the paper. 
The $ub_{dom}(T)$ can be computed in time $O(|C_T|^2(T(f) + \log |C_T|))$.


\paragraph{Knapsack Bound}
For a node $T=(S_T,C_T,W_T)$, we can easily compute $f(e|S_T)$ for any $e\in C_T$. 
Then, the optimal solution to the following integer program KNAPSACK-IP is a valid upper bound of $T$.
\begin{align*}
    \max &~ f(S_T)+ \sum_{e\in C_T} f(e|S_T) x_e  &\text{KNAPSACK-IP}\\
    s.t.& \sum_{e\in C_T} w_e x_e\le W_T      \\
    & x_e\in \{0,1\}, \forall e\in C_T 
\end{align*}
KNAPSACK-IP is clearly in the standard form of the classical \textit{knapsack problem}.
Therefore, KNAPSACK-IP is NP-Hard.
For efficiency considerations, we use the polynomial-time approximation scheme (PTAS) in \cite[Chapter 8.2]{vazirani2001approximation} to estimate an upper bound of KNAPSACK-IP, which in turn is an upper bound of the restricted SKP w.r.t. $T$.
Specifically, we round up the marginal gain of each element in $C_T$ and use dynamic programming to obtain a value that is at most $(1+\epsilon)$ times of the optimal solution of KNAPSACK-IP, where $\epsilon$ is a given parameter.
The details and proofs of this bound are left in the appended file.
We denote this upper bound as $ub_k(T)$, which can be obtained in $O(|C_T|^3 / \epsilon)$ time. 


\paragraph{Fractional Knapsack Bound}
By relaxing KNAPSACK-IP, that is, replacing $x_e\in \{0,1\}$ with $0\le x_e\le 1$, we obtained the linear relaxation problem of KNAPSACK-IP, namely the \textit{fractional knapsack problem}. The optimal solution to this relaxation problem is clearly an upper bound of KNAPSACK-IP. This bound is computed in the following simple way. 
\begin{itemize}
    \item Sort elements in $C_T$ in non-increasing order of $f(e|S_T) / w_e$.
    \item Suppose that the elements are sorted as $e_1,...,e_n$  where $n=|C_T|$. Find the largest $l$ where $w(\{e_1,...,e_l\})\le W_T$.
    \item  If $l = n$, return $f(S_T)+\sum_{j=1}^{n}f(e_j|S_T)$, otherwise return $f(S_T)+\sum_{j=1}^{l}f(e_j|S_T) + (W_T - w(\{e_1,...,e_l\}))\frac{f(e_{l+1}|S_T)}{w_{l+1}}$.
\end{itemize}

We call this upper bound the \textit{fractional knapsack bound}, denoted as $ub_{fk}(T)$. 
It can be computed in $O(|C_T| \log |C_T|)$ time. 
In particular, if the unit gains of all elements are precomputed and sorted, then $ub_{fk}$ can be computed in $O(\log |C_T|)$ time.



% ...

\paragraph{Refined Subset Bound}
Now, we introduce a novel \textit{refined subset bound}.
Given a node $T = (S_T, C_T, W_T)$ and a set $X\subseteq C_T$, let us further denote $f(X|S_T)$ as $g(X)$, and denote $f(X|S_T \cup S)$ as $g(X|S)$ for notational simplicity.
Suppose $S_T\cup X_T^*$ is an optimal solution of the restricted SKP to $T$, i.e. $X_T^*=\arg\max_{X \subseteq C_T} g(X)\ s.t.\ w(X)\leq W_T$. 
Then, we have
\begin{equation}
    \begin{aligned}
        g(X_T^*) \leq& g(X_T^* \cup X)\\
        =& g(X)+g(X_T^*|X) \\
        \leq& g(X) + \max_{S \subseteq C_T,w(S)\leq W_T} g(S|X) \\
        =& g(X) + \text{SKP}(g(\cdot|X), C_T, W_T)\\
        \leq& g(X) + ub(\text{SKP}(g(\cdot|X), C_T, W_T))
    \end{aligned}
    \label{eq:ub}
\end{equation}
where  $X\subseteq \Universe$ is an arbitrary subset, and $ub(\cdot)$ represents a valid upper bound of the SKP. The first inequality holds due to the monotonicity of $f(\cdot)$, and the second inequality holds because $X_T^*$ satisfies the constraints.

Inequality \eqref{eq:ub} generalizes the existing bound by setting $X$ as $\emptyset$. However, by selecting an appropriate $X$, we may obtain an even tighter upper bound.

Suppose $ub(\cdot)$ is equal to $ub_{fk}(\cdot)$. 
By Inequality \eqref{eq:ub}, the tightest bound of $g(X_T^*)$ can be obtained by enumerating every subset of $C_T$. This is clearly inefficient.
Rethink that we run GreedyAdd as a primal heuristic at every search node, we can reuse the subsets produced by GreedyAdd for estimating $g(X_T^*)$.
Let us use \( \mathcal{X} = (X_0, X_1, X_2, \dots, X_{|C_T|}) \) to represent the set \( X \) obtained during the execution of GreedAdd. 
Specifically, \( X_0 = \emptyset \), \( X_i \) is the set obtained at the end of the \( i \)-th iteration, and $X_{|C_T|}$ is the final set by GreedyAdd.
Then the refined subset bound is defined as follows:
\begin{align*}
ub_{rs} (T)=& f(S_T) \\
+& \min_{X_i \in \mathcal{X}} \left( g(X_i) + ub_{fk}(\text{SKP}(g(\cdot|X_i), C_T, W_T)) \right)
\end{align*}

Clearly, $ub_{rs}(T)$ is at least as good as $ub_{fk}(T)$ because \( X_0 =\emptyset\).
Since we use the fractional knapsack bound, $ub_{rs}(T)$ can be computed in time $O(|C_T|^2(T(f) + \log |C_T|))$. Specifically, for each set in $\mathcal{X}$, we need to find the element with the highest unit gain from $C_T$ in time  $O(|C_T|T(f))$, and compute fractional knapsack bound in time $O(|C_T|\log |C_T|)$.

Now, we show that the refined subset bound is actually tight.  Specifically, by the following theorem, when the weight of each element $w_i\ll W$, the reduced subset bound does not exceed $\frac{1}{1-e^{-1}} \approx 1.582$ times the optimal solution.

\begin{theorem}
In sequence $\mathcal{X}$, if there exists a $p$ such that for all $1 \leq i \leq p$, the condition $|X_{i-1}| + 1 = |X_i|$ holds. This implies that GreedyAdd can iteratively add the element with the highest unit gain from $C_T$ in each of the first $p$ iterations. Then 
\label{t:app}
$$
ub_{rs} (T) -f(S_T) \leq \dfrac{g(X_{p})}{1-e^{-\frac{w(X_{p})}{W_T}}} \leq \dfrac{g(X_T^*)}{1-e^{-\frac{w(X_{p})}{W_T}}}
$$
\end{theorem}

For convenience in later proofs, we need to recall Lemma \ref{l:wolsey} from \cite{wolsey1982maximising}.

\begin{lemma}[\cite{wolsey1982maximising}]
\label{l:wolsey}
If $P$ and $Q$ are arbitrary positive integers, $\rho_i, i=1,2,\dots,P$, are arbitrary reals, such that
$$
\min_{t=1,\dots,P}\left(\sum_{i=1}^{t-1}\rho_i + D \rho_t \right) > 0
$$
then
$$
\frac{\sum_{i=1}^P \rho_i}{\min_{t=1,\dots,P}\left(\sum_{i=1}^{t-1}\rho_i + D \rho_t \right)} \geq 1-\left( 1-\frac{1}{D}\right)^P > 1- e^{-\frac{P}{D}}
$$
\end{lemma}


\begin{proof}
Let $e_i$ denote the element added to \( X \) in the \( i \)-th iteration, where \( i \in \{1, 2, \dots, p\} \), and its marginal gain is \( \delta_i = f(e_i | S_T \cup \{e_1, e_2, \dots, e_{i-1}\}) = g(e_i|X_{i-1}) \).

Although the weight \( w_{e_i} \) of each element is a real number, we can transform it into an integer by multiplying it by a sufficiently large constant \( N \), resulting in \( w'_{e_i} = N w_{e_i} \).
At this point, the unit gain of each element is given by \( u_{e_i} = \delta_{e_i} / w'_{e_i} \).

We have obtained a sequence consisting of \( w'_{e_i} \) units of \( u_{e_i} \):
$$
\underbrace{u_{e_1},\dots,u_{e_1}}_{w'_{e_1}},\underbrace{u_{e_2},\dots,u_{e_2}}_{w'_{e_2}},\dots, \underbrace{u_{e_{p}},\dots,u_{e_{p}}}_{w'_{e_{p}}}
$$

The length of this sequence is $Nw(X_{p})$. By Lemma \ref{l:wolsey}, we can derive the following inequality:
\begin{equation}
\begin{aligned}
&\frac{\sum_{i=1}^{Nw(X_{p})} \rho_i}{\min_{t=1,2,\dots,Nw(X_{p})}\left(\sum_{i=1}^{t-1}\rho_i + NW_T \rho_t \right)} \\
\geq &1-\left( 1-\frac{1}{NW_T}\right)^{Nw(X_{p})}\\
> &1- e^{-\frac{w(X_{p})}{W_T}}
\end{aligned}
\label{eq:app}
\end{equation}

Furthermore, we have
\begin{equation}
\sum_{i=1}^{Nw(X_{p})} \rho_i = g(X_{p}) \leq g(X_T^*)
\label{eq:upper}
\end{equation}
where the inequality holds because $S_T\cup X_T^*$ is the optimal solution to node $T$.

On the other hand, 
\begin{equation}
\begin{aligned}
    &\min_{t=1,2,\dots,Nw(X_{p})}\left(\sum_{i=1}^{t-1}\rho_i + NW_T \rho_t \right) \\
    =&\min_{t=1,2,\dots,p}\left(\sum_{i=1}^{t-1}w_{e_i}u_{e_i} + NW_T u_{e_t} \right) \\
    =&\min_{t=1,2,\dots,p}\left(g(X_{t-1}) + W_T \frac{\delta_{e_t}}{w_{e_t}} \right) \\
    \geq& \min_{t=1,2,\dots,p}\left(g(X_{t-1}) + ub_{fk}(\text{SKP}(g(\cdot|X_{t-1}), C_T, W_T) \right) \\
    \geq& ub_{rs}(T) - f(S_T)
\end{aligned}
\label{eq:lower}
\end{equation}
where the first inequality holds because the unit gain of each item selected by $ub_{fk}$ is not greater than $\delta_{e_i} / w_{e_i}$, and the second inequality is from the definition of $ub_{rs}(T)$.

By combining Inequalities \eqref{eq:app}, \eqref{eq:upper} and \eqref{eq:lower}, we can obtain the final result:
$$
ub_{rs} (T) -f(S_T) \leq \dfrac{g(X_{p})}{1-e^{-\frac{w(X_{p})}{W_T}}} \leq \dfrac{g(X_T^*)}{1-e^{-\frac{w(X_{p})}{W_T}}}
$$
\end{proof}

When the weight of each element \( w_e = 1 \), the knapsack constraint degenerates into a cardinality constraint. In this case, the maximum feasible $p$ implies $w(X_p) = W_T$, $X_p=X_{|C_T|}$ and
\[
ub_{rs}(T) - f(S_T) \leq \frac{g(X_{|C_T|})}{1 - e^{-1}}
\]

This means that for the submodular function maximization problem under the cardinality constraint, $ub_{rs}$ is tighter than the bound obtained by dividing the objective value of the greedy algorithm by $(1-e^{-1})$.


\section{The Branching Method}

\begin{figure*}[!t]
  \centering
  \begin{subfigure}{0.48\textwidth}
    \centering
    \includegraphics[height=4.5cm]{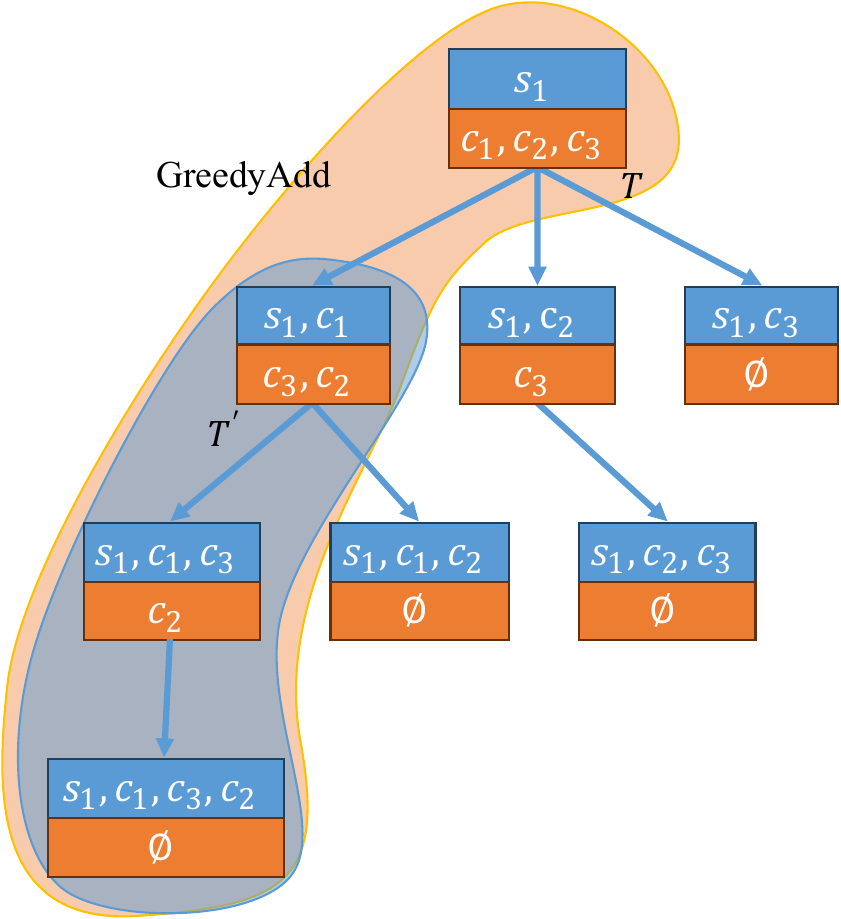}
    \caption{Basic Branching Method}
    \label{fig:basic}
  \end{subfigure}
  \hfill
  \begin{subfigure}{0.48\textwidth}
    \centering
    \includegraphics[height=3.2cm]{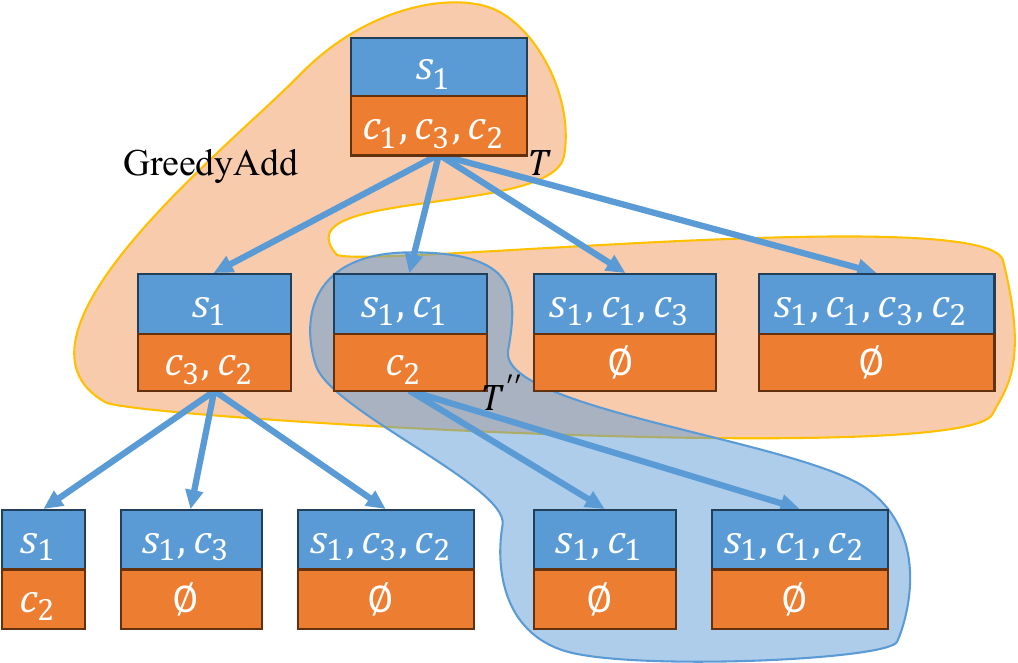}
    \caption{Dual Branching Method}
    \label{fig:dual}
  \end{subfigure}
  \vspace{\baselineskip}
  \caption{The branching at $T = (\{s_1\}, \{c_1,c_2,c_3\}, W_T)$. We assume that doing GreedyAdd on $T$ successively add elements $c_1,c_3,c_2$. The GreedyAdd algorithm computes marginal gains w.r.t. the selected sets within colored zones for all elements.}
  \label{fig:branch}
\end{figure*}

Given a node \( T = (S_T, C_T, W_T) \), the branching method generates some child nodes.
The results obtained from these child nodes should be summarized as the solution to the parent node, i.e. \( \text{SKP}(f(S_T)+f(\cdot|S_T),C_T,W_T) \), without loss of optimality.

For a node \( T = (S_T, C_T, W_T) \), we denote $n = |C_T|$. For an order of elements in $C_T$, say \( c_1, c_2, \dots, c_{n} \), we denote $C_{\leq i}=\{c_1,\dots,c_i\}$. 

\paragraph{Basic Branching Method}

We sort $C_T$ by non-increasing order of the unit gains $f(c_i|S_T)/w_{c_i}$, that is, $f(c_i|S_T)/w_{c_i}\ge f(c_{i+1}|S_T)/w_{c_{i+1}}$ for any $i = 1,...,n-1$, and obtain an order of $C_T$. The basic branching method generates $n$ child nodes as shown in the following based on this order.
\begin{itemize}
    \item $T_1=(S_T\cup \{c_1\},C_T \setminus C_{\leq 1}, W_T - w_{c_1})$,
    \item $T_2=(S_T\cup \{c_2\},C_T \setminus C_{\leq 2}, W_T - w_{c_2})$,
    \item ...
    \item $T_n=(S_T\cup \{c_n\},C_T \setminus C_{\leq n}, W_T - w_{c_n})$.
\end{itemize}
Because high-unit-value elements are moved to $S_T$ with priority, this ordering heuristic allows finding high-quality solutions at an early stage.

Also, the child nodes are visited in the depth-first search. If there exists a $i$ that satisfies 
$$
f(S_T) +ub_{fk}(SKP(f(\cdot|S_T), C_T \setminus C_{\leq i}, W_T) \leq lb^*
$$
then only the first $i$ child nodes need to be generated and visited recursively.



\paragraph{Dual Branching Method}

Let \( \hat{X} \) be the elements selected by GreedyAdd at $T$.
Then, we order the elements of $C_T$ like this -- The first \( |\hat{X}| \) elements of $C_T$ are from \( \hat{X} \), sorted according to the order in which they are selected by the greedy primal heuristic, and the remaining elements in $C_T\setminus \hat{X}$ are ordered arbitrarily afterward. 
Then, the dual branching method generates $n+1$ child nodes as follows. 
\begin{itemize}
    \item $T_0=(S_T,C_T \setminus C_{\leq 1}, W_T)$,
    \item $T_1=(S_T\cup C_{\leq 1},C_T \setminus C_{\leq 2}, W_T - w(C_{\leq 1}))$,
    \item $T_2=(S_T\cup C_{\leq 2},C_T \setminus C_{\leq 3}, W_T - w(C_{\leq 2}))$,
    \item ...
    \item $T_{n-1}=(S_T\cup C_{\leq n-1},C_T \setminus C_{\leq n}, W_T - w(C_{\leq n-1})))$,
    \item $T_n=(S_T\cup C_{\leq n},C_T \setminus C_{\leq n}, W_T - w(C_{\leq n})))$.
\end{itemize}
However, for any \( i > |\hat{X}| \), the weight \( W_{T_i} \) of the child node \( T_i \) satisfies \( W_{T_i} \leq 0 \), meaning these nodes can be pruned. Therefore,  we can actually construct \( |\hat{X}| + 1 \) child nodes, probably smaller than $n+1$.
And in the dual branching method, if there exists a $i$ that satisfies
$$
f(S_{T_i}) +
ub_{fk}(SKP(f(\cdot|S_{T_i}), C_T \setminus C_{\leq i}, W_{T} - w(C_{\leq i})) \leq lb^*
$$
then only the first $i$ child nodes need to be generated and searched recursively.

\paragraph{Comparing the Basic and Dual Branching Methods}


        
    
    
    

The dual branching method can reduce some repeat computations compared with the basic branching method because we can combine the branching process and the greedy primal heuristic.
For example, in Figure \ref{fig:branch}, \( \{s_1,c_1\} \), \( \{s_1,c_1,c_3\} \) and \( \{s_1,c_1,c_3,c_2\} \) are the selected sets that would be generated during the GreedyAdd at node \( T \). To find the item with the maximum unit gain, it is necessary to compute \( f(c_i | \{s_1\}) \), \( f(c_i | \{s_1, c_1\}) \) and \( f(c_i | \{s_1, c_1, c_3\})\) for each element $c_i$. Similarly, we need to compute  \( f(c_i | \{s_1, c_1\}) \) and \( f(c_i | \{s_1, c_1, c_3\})\)  again for each element $c_i$ at node $T'$.
Such repeat computation does not occur in the dual branching method.

On the other hand,  the space complexity of dual branching  is higher than the basic branchings.
Specifically, before the greedy primal heuristic terminates, the algorithm using the dual branching method must store all the generated child nodes. As a result, at most $O(|\Universe|^2)$ search nodes need to be stored in memory. But the basic branching method only needs to store $O(|\Universe|)$ search nodes.



\section{Implementation Details}

We use $\{T_0, T_1, \dots, T_n\}$ to denote the child nodes of node $T$.

\paragraph{Lazy Update} 
The lazy update technique was early proposed in \cite{minoux2005accelerated} to speed up greedy submodular function maximization.
We incorporate this idea in our branch-and-bound algorithm.
In general, in a child node $T_i$, we do not update the marginal gains \( f(e|S_{T_i}) \) for each element $e\in C_{T_i}$; instead, we only update elements with unit gains greater than $\frac{lb^*-f(S_{T_i})}{W_{T_i}}$, while reusing the marginal gains from their parent nodes for all other elements.
Due to the submodularity of $f$, the marginal gain from the parent node is always an upper bound on the \( f(e|S_T) \) at the current node.
This technique was also used in \cite{woydt2024submodst} for exactly solving the submodular function maximization problem under cardinality constraint.



\paragraph{Reduction Method}
We also apply some reduction techniques to exclude unfruitful elements at a search node $T=(S_T, C_T, W_T)$.
\begin{enumerate}
    \item If there exists an element $e\in C_T$ such that $w_e > W_T$ or $f(e|S_T) =0$, then $e$ can be discarded from the candidate set $C_T$.
    \item If there exists an element $e\in C_T$ such that
    \begin{align*}
    f(S_T)&+f(e|S_T)\\ &+ub_{fk}(SKP(f(\cdot|S_T), C_T,W_T-w_e)) \leq lb^*
    \end{align*}
    then $e$ can be discarded from the candidate set $C_T$, as including it to $S_T$ cannot improve the current best solution.
\end{enumerate}

\section{Experiments}

\paragraph{Settings}

In this section, we empirically evaluate the proposed algorithms. 
All experiments are conducted on a machine with an Intel(R) Xeon(R) Platinum 8360Y CPU @ 2.40GHz and an Ubuntu 22.04 operating system.

\paragraph{Implemented Algorithms}
Because we discuss different branching and upper bounding methods, we implemented five branch-and-bound algorithms, \textbf{basic-dom}, \textbf{basic-k}, \textbf{basic-fk}, \textbf{basic-rs}, and \textbf{dual-rs}.
The details of each algorithm are presented in Table \ref{tab:alg}. We set the parameter $\epsilon$ in \textbf{basic-k} to $1$.
Our preliminary experiments showed that removing the primal heuristic makes \textbf{basic-k} and \textbf{basic-fk} more efficient in all test cases, although the quality of the lower bound is sacrificed a bit at some nodes.
Meanwhile, the dual branching method with $ub_{rs}$ consistently outperformed dual branching with other bounds. Hence, we believe the five variants are representative enough.

\begin{table}[]
\centering
\caption{Introduction to the implementation of each algorithm. $n$ denotes $|C_T|$.}
\label{tab:alg}
\begin{tabular}{ccccc}
\toprule
Algorithm & Primal Heur. & Bound & Branch & Time on Node \\
\midrule
basic-k   &  / & $ub_k$ & basic  &     $O(nT(f)+n^3/\epsilon)$     \\
basic-fk  &  / & $ub_{fk}$ & basic  &     $O(n(T(f)+\log n))$     \\
basic-dom &  GreedyAdd & $ub_{dom}$ & basic  &   $O(n^2(T(f) + \log n)$       \\
basic-rs  &  GreedyAdd & $ub_{rs}$ & basic  &    $O(n^2(T(f) + \log n))$      \\
dual-rs   &  GreedyAdd & $ub_{rs}$ & dual  &     $O(n^2(T(f) + \log n))$     \\
\bottomrule
\end{tabular}
\end{table}


We also test the best-known existing exact algorithm for SKP, i.e. \textbf{BFSTC} \cite{sakaue2018accelerated} and \textbf{ILP} \cite{nemhauser1981maximizing}.
Because we did not obtain the code from \cite{sakaue2018accelerated}, \textbf{BFSTC} is a re-implementation of their best $A^*$ algorithm.
The preliminary tests showed that our re-implementation is at least $3\times$ faster than their reported results for the same benchmark problems.
We believe that this re-implementation is valid because the tested machines have approximately the same CPU frequency. 
We also implement the constraint generation algorithm with the integer linear program (\textbf{ILP}) in \cite{nemhauser1981maximizing} with Gurobi 12.01. 

All algorithms are written in C++23 and compiled by gcc version 14.2.0. We use the -O3 flag in the code compilation. Multi-threading is turned off for all algorithms. Each algorithm runs for a maximum of 30 minutes for each input instance.

\begin{figure*}[!t]
  \centering
  \includegraphics[width=\textwidth]{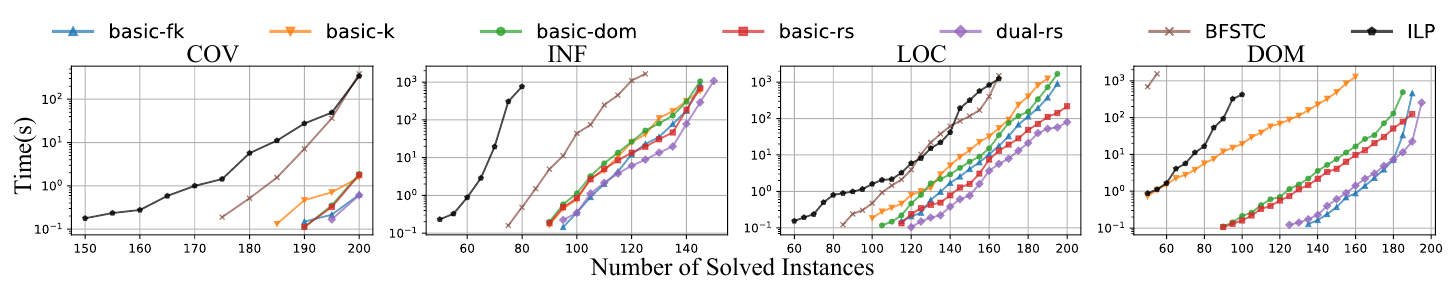}
  \caption{Number of instances solved within time limits under the normal generation method.} \label{fig:part_solved_time}
\end{figure*}

\begin{figure*}[!tbp]
  \centering
  \includegraphics[width=\textwidth]{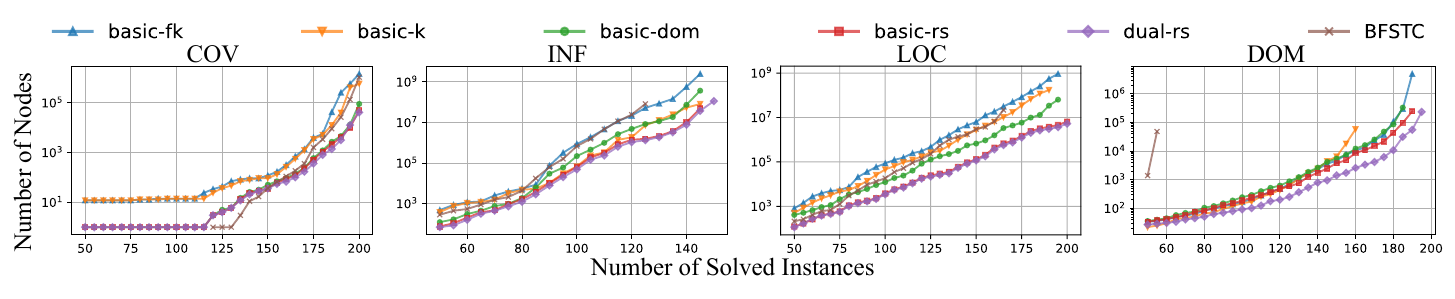}
  \caption{Number of instances solved within visited nodes limits 
  under the normal generation method.} \label{fig:part_solved_nodes}
\end{figure*}

\begin{table*}[!t]
\centering
\caption{The maximum solved $W$ value for each algorithm and their respective worst-case running times within the range $[1, \text{the maximum solved $W$}]$ are reported.
For each submodular function, we highlight the two instances with the longest execution times from our full experimental results. The complete experimental results can be found in the appended file.}
\label{tab:time_results}
\begin{tabular}{llcccccccccccccc}
\toprule
                     & \multirow{2}{*}{Instance} & \multicolumn{2}{c}{BFSTC} & \multicolumn{2}{c}{ILP} & \multicolumn{2}{c}{basic-k} & \multicolumn{2}{c}{basic-fk} & \multicolumn{2}{c}{basic-dom} & \multicolumn{2}{c}{basic-rs} & \multicolumn{2}{c}{dual-rs} \\
                     &                           & $W$           & Time(s)   & $W$          & Time(s)  & $W$            & Time(s)    & $W$          & Time(s)       & $W$             & Time(s)     & $W$             & Time(s)    & $W$         & Time(s)       \\
\midrule
\multirow{2}{*}{COV} & C.100.5.1                 & \textbf{20}   & 346       & \textbf{20}  & 342      & \textbf{20}    & 2          & \textbf{20}  & \textbf{1}    & \textbf{20}     & 2           & \textbf{20}     & 2          & \textbf{20} & \textbf{1}    \\
                     & C.100.8.1                 & \textbf{20}   & 385       & \textbf{20}  & 172      & \textbf{20}    & 1          & \textbf{20}  & \textbf{0}    & \textbf{20}     & 1           & \textbf{20}     & 1          & \textbf{20} & \textbf{0}    \\
\midrule
\multirow{2}{*}{INF} & inf\_100\_5\_1               & 7             & 385       & 4            & 1521     & 9              & 290        & 9            & 336           & 9               & 637         & 9               & 336        & \textbf{10} & \textbf{1074} \\
                     & inf\_100\_8\_1               & 7             & 949       & 3            & 19       & 10             & 1633       & 9            & 172           & 9               & 408         & \textbf{10}     & 1794       & \textbf{10} & \textbf{724}  \\
\midrule
\multirow{2}{*}{LOC} & L.60.5.1                  & 10            & 1535      & 8            & 1668     & 15             & 1251       & 19           & 1617          & 17              & 1473        & \textbf{20}     & 142        & \textbf{20} & \textbf{69}   \\
                     & L.60.8.1                  & 11            & 1736      & 10           & 1800     & 15             & 1035       & 19           & 1361          & 18              & 1684        & \textbf{20}     & 217        & \textbf{20} & \textbf{80}   \\
\midrule
\multirow{2}{*}{DOM} & ca-AstroPh                & 2             & 34        & 4            & 828      & 9              & 1740       & \textbf{18}  & \textbf{1244} & 15              & 1649        & 15              & 1619       & 17          & 1596          \\
                     & econ-orani678             & 3             & 533       & 13           & 398      & 9              & 922        & 15           & 1082          & 13              & 1185        & 19              & 113        & \textbf{20} & \textbf{19}  \\
\bottomrule
\end{tabular}
\end{table*}

\paragraph{Datasets}
 
To evaluate the algorithms, we perform experiments on the following four problems:

\begin{enumerate}[label=\arabic*.]
    \item \textbf{Weighted Coverage (COV).} We are given a set of items \( M = \{1, \dots, m\} \) and a collection of item sets \( \mathcal{C} = \{C_1, \dots, C_n\} \). The objective is to select a subset \( S \subseteq \mathcal{C} \) such that the union of the selected sets maximizes the total value. Specifically let each item \( j \in M \) have a value \( v_j \). The objective to maximize is given by:
    \begin{equation*}
    f(S) = \sum_{i \in \bigcup_{j \in S} C_j} v_j 
    \end{equation*}

    \item \textbf{Bipartite Influence (INF).} Let $\Universe = \{1,\dots,n\}$ be a set of sources and $M=\{1,\dots,m\}$ be a set of targets. Consider a bipartite graph \( G = (\Universe \cup M, E) \), where each edge \( (j, i) \) is associated with a weight \( 0 \leq p_{ij} \leq 1 \), representing the activation probability of source \( j \) influencing target \( i \). If the edge \( (j, i) \) is not present in \( G \), then \( p_{ij} = 0 \). The probability that a target \( i \) is activated by a set of sources \( S \subseteq \Universe \) is given by $1 - \prod_{j \in S} (1 - p_{ij})$.
    Thus, the objective to maximize is:
    \begin{equation*}
        f(S)=\sum_{i \in M}\left( 1 - \prod_{j \in S} (1 - p_{ij}) \right)
    \end{equation*}

    \item \textbf{Facility Location (LOC).} \cite{uematsu2020efficient} described this variant of the facility location problem. Let $\Universe = \{1,\dots,n\}$ be a set of factory locations and $M=\{1,\dots,m\}$ be a set of customers. The value \( v_{ij} \) represents the profit that customer \( i \in M \) can obtain from factory \( j \in \Universe \). For a subset $S \subseteq \Universe$ of selected facilities, each customer is assigned to the facility that provides the highest profit. The objective to maximize is:
    \begin{equation*}
    f(S)=\sum_{i \in M} \max_{j \in S} v_{ij}
    \end{equation*}


    \item \textbf{Partial Dominating Set (DOM).} 
    Given an undirected graph $G=(\Universe, E)$, Let $N(v)$ represent the set of all neighbors of vertex $v$, and define $N[v] = N(v)\cup \{v\}$. The objective is to select a subset $S\subseteq \Universe$ such that the number of dominated vertices is maximized. The objective to maximize is:
    \begin{equation*}
    f(S)=\left| \bigcup_{v\in S} N[v] \right|
    \end{equation*}
    
\end{enumerate}

Like \cite{woydt2024submodst}, we sample 10 same benchmark data for each of the COV, INF, LOC and DOM problems. 
The data of COV and LOC are sampled from the dataset in \cite{uematsu2020efficient}, while the data of INF are from \cite{csokas2024constraint}. 
For each data of these three problems, the values of $W$ are set to $1, 2, ..., 20$. 
For the DOM problem, we use the largest connected subgraphs of 10 graphs from the Network Repository \cite{rossi2015network}. 
Likewise, the values of $W$ are also set to $1, 2,\ldots, 20$.
Hence, for each of these four problems, we test $200$ instances.


In the original datasets, no weight is assigned to each element. Here, we generate the weights $w_e$ as follows.
\begin{enumerate}
    \item \textbf{Normal}: The weights are generated from a normal distribution \( \mathcal{N}(1, 0.2) \) and then clamped to the range $[0.1, 1.9]$.
    \item \textbf{Uniform}: The weights are generated from a uniform distribution $\mathcal{U}(0.4, 1.6)$.
    \item \textbf{Unit}: The weight of any elements is 1.
\end{enumerate}
Each instance uses 0 as a random seed.

\begin{figure*}[!t]
    \centering
    \includegraphics[width=\textwidth]{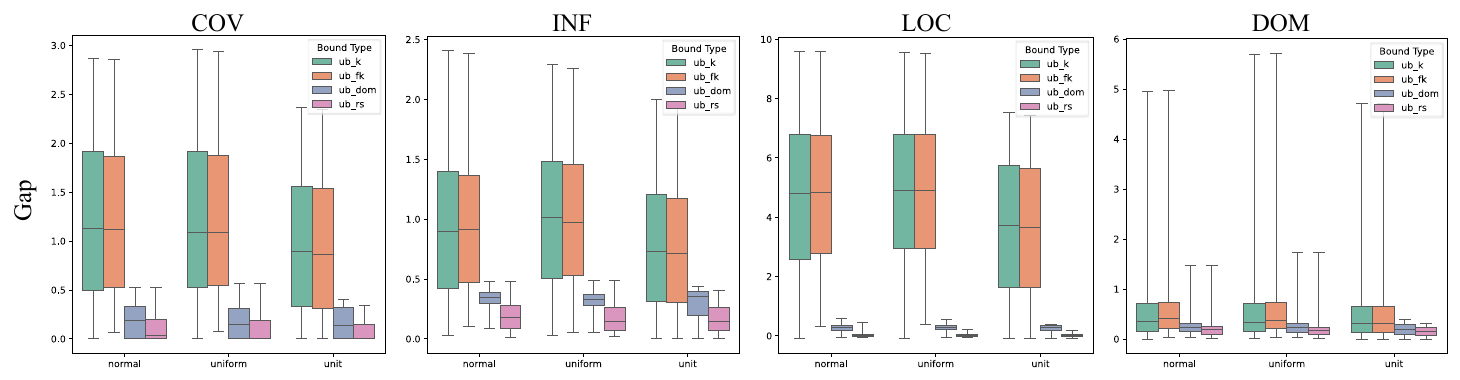}
    \caption{Gap values $\frac{ub-s^*}{s^*}$ of different upper bounds under various conditions. } \label{fig:gap_box}
\end{figure*}

\begin{table*}[]
\centering
\caption{Comparison with ICG \cite{uematsu2020efficient}, ICG$(k-1)$, GCG, ECG \cite{csokas2024constraint} and SubModST \cite{woydt2024submodst}. Running time is measured in seconds (s).}
\label{tab:Cardinality}
\begin{tabular}{llcccccccc}
\toprule
                     & Instance & $|U|$ & $W$ & ICG                & ICG$(k-1)$           & GCG                & ECG                & SubModST & dual-rs \\
\midrule
\multirow{2}{*}{COV} & C.100.5.1                    & 100                   & 5                     & \textgreater{}7200 & 676.98             & 377.66             & 756.57             & \textbf{0.01}                         & \textbf{0.01}                        \\
                     & C.100.8.1                    & 100                   & 8                     & 2797.19            & 4166.53            & 2054.98            & 5078.67            & \textbf{0.10}                         & 0.50                        \\
\midrule
\multirow{4}{*}{INF} & I.100.5.1                    & 100                   & 5                     & \textgreater{}7200 & \textgreater{}7200 & \textgreater{}7200 & 639.71             & \textbf{0.01}                         & \textbf{0.01}                        \\
                     & inf\_100\_5\_1                    & 100                   & 11                    & /                  & /                  & /                  & /                  & \textbf{589.31}                       & 803.55                      \\
                     & I.100.8.1                    & 100                   & 8                     & \textgreater{}7200 & \textgreater{}7200 & \textgreater{}7200 & \textgreater{}7200 & \textbf{0.01}                         & \textbf{0.01}                        \\
                     & inf\_100\_8\_1                    & 100                   & 11                    & /                  & /                  & /                  & /                  & \textbf{604.53}                       & 796.83                      \\
\midrule
\multirow{4}{*}{LOC} & L.60.5.1                     & 60                    & 5                     & \textgreater{}7200 & 6016.81            & 4988.19            & 3217.13            & \textbf{0.01}                         & 0.05                        \\
                     & L.60.5.1                     & 60                    & 20                    & /                  & /                  & /                  & /                  & 1307.38                      & \textbf{81.92}                       \\
                     & L.60.8.1                     & 60                    & 8                     & \textgreater{}7200 & \textgreater{}7200 & \textgreater{}7200 & \textgreater{}7200 & \textbf{0.28}                         & 0.69                        \\
                     & L.60.8.1                     & 60                    & 20                    & /                  & /                  & /                  & /                  & 1002.74                      & \textbf{34.67}                       \\
\midrule
\multirow{2}{*}{DOM} & ca-AstroPh                & 17903                 & 20                    & /                  & /                  & /                  & /                  & \textbf{249.58}                       &           1161.09                  \\
                     & econ-orani678              & 2529                  & 20                    & /                  & /                  & /                  & /                  & \textgreater{}1800           & \textbf{1.40}                       \\
\bottomrule
\end{tabular}
\end{table*}


\paragraph{The Running Time}

In Figure \ref{fig:part_solved_time} and Table \ref{tab:time_results}, we show the experimental results where the element weights follow normal distributions. 
The experimental results show that \textbf{basic-fk} exhibits the best performance in all algorithms using the basic branching method. And \textbf{dual-rs} shows the best overall performance.
Examinations of \textbf{basic-rs} and \textbf{dual-rs} show that the dual branching method achieves a 
 \textasciitilde$2\times$ acceleration in algorithm execution, especially for DOM problems.



\paragraph{The Number of Nodes}

We further compare the number of nodes searched by each algorithm. 
Figure \ref{fig:part_solved_nodes} shows the number of instances solved by each algorithm against a specified number of nodes; still, element weights follow normal distributions. 
we noticed that the \textbf{basic-rs} generates fewer nodes for all kinds of problems. 
Additionally, the \textbf{dual-rs} approach visits fewer nodes compared to \textbf{basic-rs}, with a notably large decrease for the DOM problem.

Note that comparable conclusions in terms of running time and number of nodes can be drawn for instances of other weight distributions; therefore, these results have been included in the appended file.

\paragraph{The Tightness of Upper Bound}
Figure \ref{fig:gap_box} shows the gap \footnote{For a problem with optimal objective $s^*$ and an upper bound $ub$, we define \textit{gap} between $s^*$ and $ub$ as $\frac{ub-s^*}{s^*}$.} between optimal objective and upper bounds in instances where optimal solutions are known.

Clearly, $ub_{dom}$ is tighter than both $ub_k$ and $ub_{fk}$ in various problems. 
The novel $ub_{rs}$ remains the tightest in all tested instances.

\paragraph{The SKP under Cardinality Constraint}
When the weight distribution is Unit, the knapsack constraint degenerates into a cardinality constraint.
For instance of these cases, there are more refined existing algorithms including 
\textbf{ICG} \cite{uematsu2020efficient}, \textbf{ICG}$(k-1)$, \textbf{GCG}, \textbf{ECG} \cite{csokas2024constraint} and \textbf{SubModST} \cite{woydt2024submodst}. 
The codes of \textbf{ICG}, \textbf{ICG}$(k-1)$, \textbf{GCG}, \textbf{ECG} are not publicly available but the codes of \textbf{SubModST} are fully open-source. We sample the results of ILP-based algorithms from \cite{csokas2024constraint}.

Tables \ref{tab:Cardinality} present the experimental results.
They show that the combinatorial \textbf{dual-rs} and \textbf{SubModST} currently outperform other algorithms, which are based on ILP solvers. 
The \textbf{dual-rs} and \textbf{SubModST} compete with each other for different instances. 
Notably, in DOM cases like econ-orani678, \textbf{dual-rs} achieves a significant speedup over \textbf{SubModST}.

\section{Conclusion and Discussion}
In this paper, we provide a branch-and-bound algorithm and a new upper bound for the SKP, called the refined subset bound.
Experiments show that our algorithm significantly outperforms existing exact algorithms. Meanwhile, the proposed refined subset bound is tighter than the existing upper bounds.
Additionally, we use the dual branching method to minimize repeat computation and further reduce the total number of search nodes.

In the future, we can explore submodular maximization problems under more general constraints, such as matroid constraints.
On the other hand, it is interesting to investigate exact parallel algorithms for SKP.




\bibliography{mybibfile}

\end{document}